\def\rfr#1{eq. (\ref{#1})}

\def\dert#1#2{\frac{{{d}}{#1}}{{{d}}{#2}}}              

\def\bar{\begin{eqnarray}}
\def\ear{\end{eqnarray}}
\def\bb{\bibitem}
\def\eqi{\begin{equation}}
\def\eqf{\end{equation}}
\def\eqia{\begin{eqnarray}}
\def\eqfa{\end{eqnarray}}
\def\rp#1#2{{#1\over#2}}

\def\lb#1{\label{#1}}

\def\oc2{$\mathcal{O}(c^{-2})$}

\def\WD{WD0137-349}

\documentclass[11pt]{article}
\usepackage{amsmath,amsthm,amscd,amssymb}
\usepackage{latexsym}
\usepackage{graphicx,epsfig}
\usepackage{natbib}
\usepackage{ifpdf}
\ifpdf
\usepackage[%
  pdftitle={Instructions for use of the document class
    elsart},%
  pdfauthor={Simon Pepping},%
  pdfsubject={The preprint document class elsart},%
  pdfkeywords={instructions for use, elsart, document class},%
  pdfstartview=FitH,%
  bookmarks=true,%
  bookmarksopen=true,%
  breaklinks=true,%
  colorlinks=true,%
  linkcolor=red,anchorcolor=red,%
  citecolor=blue,filecolor=red,%
  menucolor=red,pagecolor=red,%
  urlcolor=cyan]{hyperref}
\else
\usepackage[%
  breaklinks=true,%
  colorlinks=true,%
  linkcolor=red,anchorcolor=red,%
  citecolor=blue,filecolor=red,%
  menucolor=red,pagecolor=red,%
  urlcolor=cyan]{hyperref}
\fi

\begin{document}

\noindent{\bf \LARGE{Dynamical constraints on some orbital and physical properties of the \WD A/B binary system}}
\\
\\
\\
{Lorenzo Iorio}\\
{\it Viale Unit$\grave{a}$ di Italia 68, 70125\\Bari, Italy
\\tel./fax 0039 080 5443144
\\e-mail: lorenzo.iorio@libero.it}

\begin{abstract}
In this paper I deal with the \WD\ binary system consisting of a white dwarf (WD) and a brown dwarf (BD) in a close circular orbit of about 116 min. I, first, constrain the admissible range of values for the inclination $i$ by noting that, from looking for deviations from the third Kepler law,  the quadrupole mass moment $Q$ would assume unlikely large values, incompatible with zero at more than 1-sigma level for $i\lesssim 35$ deg and $i \gtrsim 43$ deg. Then, by conservatively assuming that the most likely values for $i$ are those that prevent such an anomalous behavior of $Q$, i.e. those for which the third Kepler law is an adequate modeling of the orbital period, I obtain $i=39\pm 2$ deg. Such a  result  is incompatible with the value $i=35$ deg quoted in literature by more than 2 sigma. Conversely, it is shown that the white dwarf's mass range obtained from spectroscopic measurements  is compatible with my experimental range, but not for $i=35$ deg. 
As a consequence, my estimate of $i$ yields an orbital separation of $a=(0.59\pm 0.05)$R$_{\odot}$ and an equilibrium temperature of BD of $T_{\rm eq}=(2087\pm 154)$K which differ by $10\%$ and $4\%$, respectively, from the corresponding values for $i=35$ deg.
 \end{abstract}

 \section{Introduction}
 In general, binary systems composed by a white dwarf (WD) orbited by a brown dwarf (BD) as a companion are rare:   the \WD\ (BPS CS 29504-0036) system \citep{Max06}, consisting of a WD and a  BD orbiting in a 116 min close circular path, belongs to such a class.  BD must have survived a previous phase in which it was engulfed by the red giant progenitor of WD \citep{Poli04} experiencing orbital drag which notably shrank its orbit, originally much wider.

 In \citep{Iorio07} a dynamical determination of its quadrupole mass moment $Q$ from deviations of the third Kepler law was claimed obtaining an unlikely large value for it: $Q=-1.5\times 10^{47}$ kg m$^2$, with $|Q/MR^2|\approx 10^3$, where $M$ and $R$ are the WD's mass and equatorial radius. In this paper I clarify the meaning of such a result and show how it can be used to constrain the inclination angle $i$. My analysis furnishes a physical  justification of the use of the third Kepler law in modeling the orbital period of this particular system, without assuming it uncritically a priori. Moreover, it also  corrects an error concerning $i$ in \citep{Max06} in which $i=35$ deg is quoted, yielding an orbital separation of 0.65R$_{\odot}$. Such values for $i$ and $a$ are reported in other works being used for investigations on various aspects of the \WD\ system like, e.g., the determination of $Q$ itself \citep{Iorio07} and the heating of BD \citep{Bur06}.
\section{Constraining the inclination with the quadrupole mass moment}
The phenomenologically measured period amounts to  \citep{Max06}
\eqi P_{\rm b} = 0.083\pm 0.0002\ {\rm d}:\eqf in principle, it accounts for all the dynamical effects affecting our binary system to the
measurement accuracy. Let us, now, contrast it to the purely Keplerian period
\eqi P^{(0)}=2\pi\sqrt{\rp{a^3}{G(M+m)}},\lb{pkep}\eqf where $a$ is the relative semimajor axis, and $M$ and $m$ are the masses of WD and BD, respectively, in order to see if it is compatible with $P_{\rm b}$, within the errors, or if some significant discrepancy occurs.   Such a comparison does, indeed, make sense because the ``ingredients" entering \rfr{pkep} have all been determined in a way which is independent of the third Kepler law itself, apart from the inclination $i$ which will, thus, be treated as a free parameter. Indeed, the mass of WD \citep{Max06},
\eqi M=(0.39\pm 0.035){\rm M}_{\odot},\lb{massaWD}\eqf   was non-dynamically inferred from its effective temperature $T_{\rm eff}$ and  surface gravity $\log g$ \citep{Driebe98, Ben99}- measured, in turn, from an analysis of the hydrogen absorption lines of the optical spectrum of WD \citep{Kos01}-using well tested and reliable models of white dwarfs. Then, from \eqi \rp{K_{m}P_{\rm b}}{2\pi}=x_{m}=\rp{M a\sin i}{(M+m)},\lb{rav}\eqf where $K_{M/m}$ are the projected semiamplitudes of the radial velocities, phenomenologically measured from the spectroscopic radial velocity curves, $x_{M/m}$ are the projected barycentric semimajor axes and $i$ is the inclination of the orbital plane to the plane of the sky, it is possible to measure the ratio of the masses $\gamma=m/M$ as     \citep{Max06}
\eqi \gamma =\rp{K_M}{K_m}=0.134\pm 0.006:\lb{gamma}\eqf I have used \citep{Max06}
\begin{equation}\left\{\begin{array}{lll}
K_{ M} = 27.9.5\pm 0.3\ {\rm km\ s}^{-1},\\\\
K_{ m} = 208.5\pm 8.1\ {\rm km\ s}^{-1}.
\lb{velrad}
\end{array}\right.\end{equation}
From the value of $M$ and \rfr{gamma} the mass of the BD follows  \citep{Max06}\eqi m=(0.052\pm 0.007){\rm M}_{\odot}.\eqf  Now, from \eqi a = \left( 1+\gamma\right)\rp{x_m}{\sin i},\lb{sam}\eqf it is possible to express $P^{(0)}$ in terms of known quantities, apart from the inclination $i$ which will be treated as an independent variable on which I want to put some dynamical constrains. Note that in \citep{Max06} the value $i=35$ deg is released yielding an orbital separation $a=0.65$R$_{\odot}$ which, in turn, was used to obtain an equilibrium temperature of BD of \citep{Bur06}
 \eqi T_{\rm eq}^{(\rm BD)}=T^{(\rm WD)}\sqrt{\rp{R}{2a}}\approx 2000{\rm K}.\eqf
 As we will see, such a value for $i$ is probably wrong.

\begin{figure}
\begin{center}
\includegraphics[width=13cm,height=11cm]{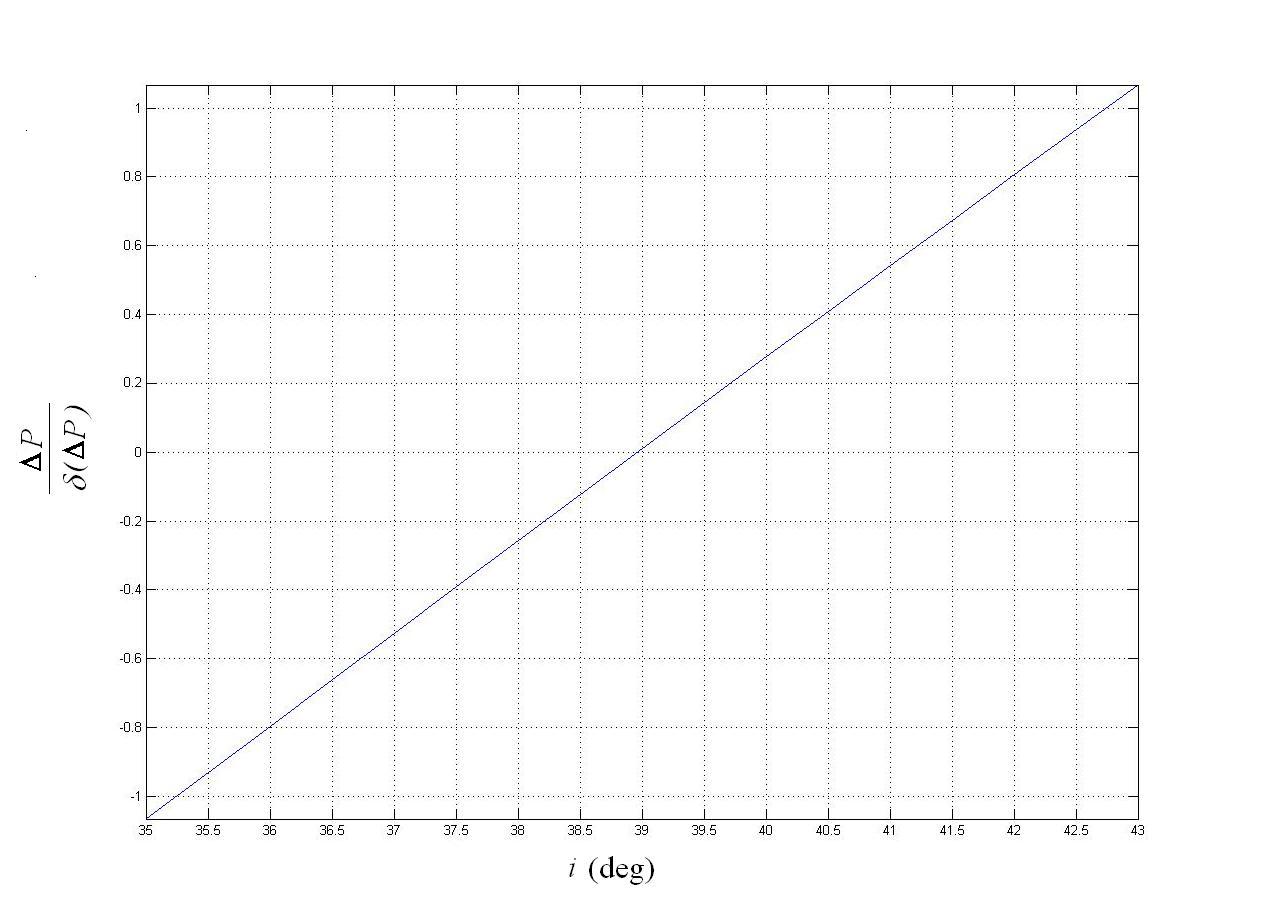}
\end{center}
\caption{\label{periodo} Discrepancy $\Delta P$ between the measured orbital period $P_{\rm b}$ and the computed Keplerian one $P^{(0)}$, normalized to its error $\delta(\Delta P)$, as function of the inclination $i$. It becomes significant, at more than 1-sigma level, for $i\lesssim 35$ deg and $i\gtrsim 43$ deg.}
\end{figure}

Let us investigate   the ratio
\eqi \rp{ \Delta P  }{ \delta(\Delta P ) }\equiv\rp{P_{\rm b}-P^{(0)}}{\delta P_{\rm b} + \delta P^{(0)}}\eqf as a function of $i$. From Figure \ref{periodo}  it can be noted  that outside the interval $35\ {\rm deg} \lesssim i \lesssim 43$ deg such a discrepancy becomes significative at more that 1-sigma level changing its sign as well: inside such a range it is, instead, compatible with zero.

May such a discrepancy have a physical meaning, so that the associated figures for $i$ can be considered realistic values for it?
Deviations from the third Kepler law could be induced by  dynamical effects neglected in modeling the orbital period. However, post-Newtonian corrections of order $\mathcal{O}(c^{-2})$ \citep{Sof89} are negligible because much smaller than the experimental error $\delta P_{\rm b}=0.0002$ d. Another possible candidate is the post-Keplerian, Newtonian effect of the quadrupole mass moment $Q$ of  WD \citep{Iorio07}. Let us explore this possibility in order to see if it is viable; such an analysis will also better clarify the strange result obtained in \citep{Iorio07} for $i=35$ deg. We will see that it can yield us useful insights about the inclination of the considered binary system.

\begin{figure}
\begin{center}
\includegraphics[width=13cm,height=11cm]{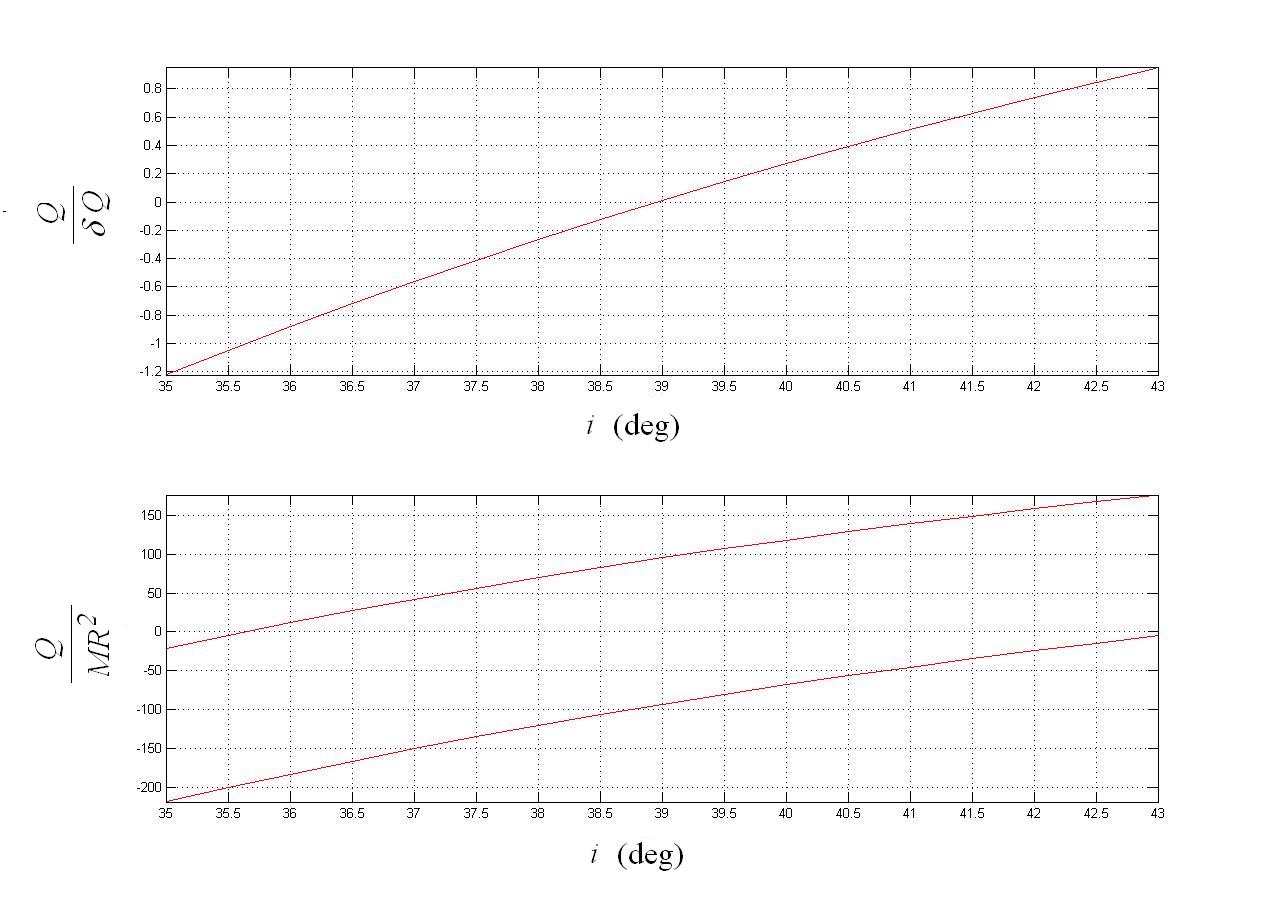}
\end{center}
\caption{\label{qu}  The adimensional quantities $Q/\delta Q$ (upper panel) and $Q/MR^2$ (lower panel) as functions of the inclination $i$. The upper curve in the lower panel is determined by $Q+\delta Q$, while the lower curve in the lower panel is determined by $Q-\delta Q$, so that the admissible values for $Q/MR^2$ lie in the region delimited by such two curves. }
\end{figure}

By using \rfr{periodoQ} of Appendix
\eqi Q =\rp{ P_{\rm b}\sqrt{Ga(M+m)^3(1-e^2)^3}
}{3\pi}-\rp{2}{3}(M+m)a^2(1-e^2)^{3/2},\lb{QU}\eqf
in Figure \ref{qu} I plot $Q/\delta Q$ and $Q/MR^2$ versus $i$, where $\delta Q$ is given by \rfr{erroreQ}-\rfr{erroriQ} of Appendix.
It can be noted that $Q$, assumed as responsible of the discrepancy between $P_{\rm b}$ and $P^{(0)}$, becomes, in fact, a determined quantity at more than 1-sigma level outside about $35 \ {\rm deg} \lesssim i \lesssim 43$ deg: WD would be highly oblate ($Q< 0$) for $i\lesssim 35$ deg and highly prolate  ($Q>0$) for $i \gtrsim 43$ deg. The physical meaning of such an outcome is, however, suspect because $Q/MR^2$, where $R=(0.0186\pm 0.0012){\rm R}_{\odot}$ is the WD's radius \citep{Max06} would get as large as 150-200. Such large values of $Q/MR^2$ are unlikely; WD's quadrupole moment (normalized to $MR^2$) are expected to be smaller than 1 \citep{Bay71, Papo}.   In principle, it could be argued that the determined $Q$ is, in fact, an ``effective'' quantity which accounts for other, unknown dynamical effects.
A more conservative interpretation of our results is that the inclination of the orbital plane of the \WD\ system is confined in that interval in which  the third Kepler law is an adequate description of the orbital dynamics and $Q$ is compatible with zero.

I  will, now, work within such a framework. Since I have information on $M$ independent of the third Kepler law itself, I can use them to further constrain $i$. Indeed, by equating $P_{\rm b}$ to $P^{(0)}$ and using \rfr{rav} and \rfr{sam} I have \eqi i = \arcsin\left\{\left[\rp{(1+\gamma)^2 K_m^3 P_{\rm b}}{2\pi GM}\right]^{1/3}\right\}=39\pm 2\ {\rm deg}.\lb{angolaz}\eqf    Such a value is incompatible with $i=35$ deg quoted in \citep{Max06} at more than 2-sigma level. The relative semimajor axis, computed with \rfr{sam} and \rfr{angolaz},  becomes
\eqi a = (0.59\pm 0.05){\rm R}_{\odot},\eqf     yielding an equilibrium temperature for BD of \eqi T_{\rm eq}^{(\rm BD)}=(2087\pm 154){\rm K}.\eqf

\begin{figure}
\begin{center}
\includegraphics[width=13cm,height=11cm]{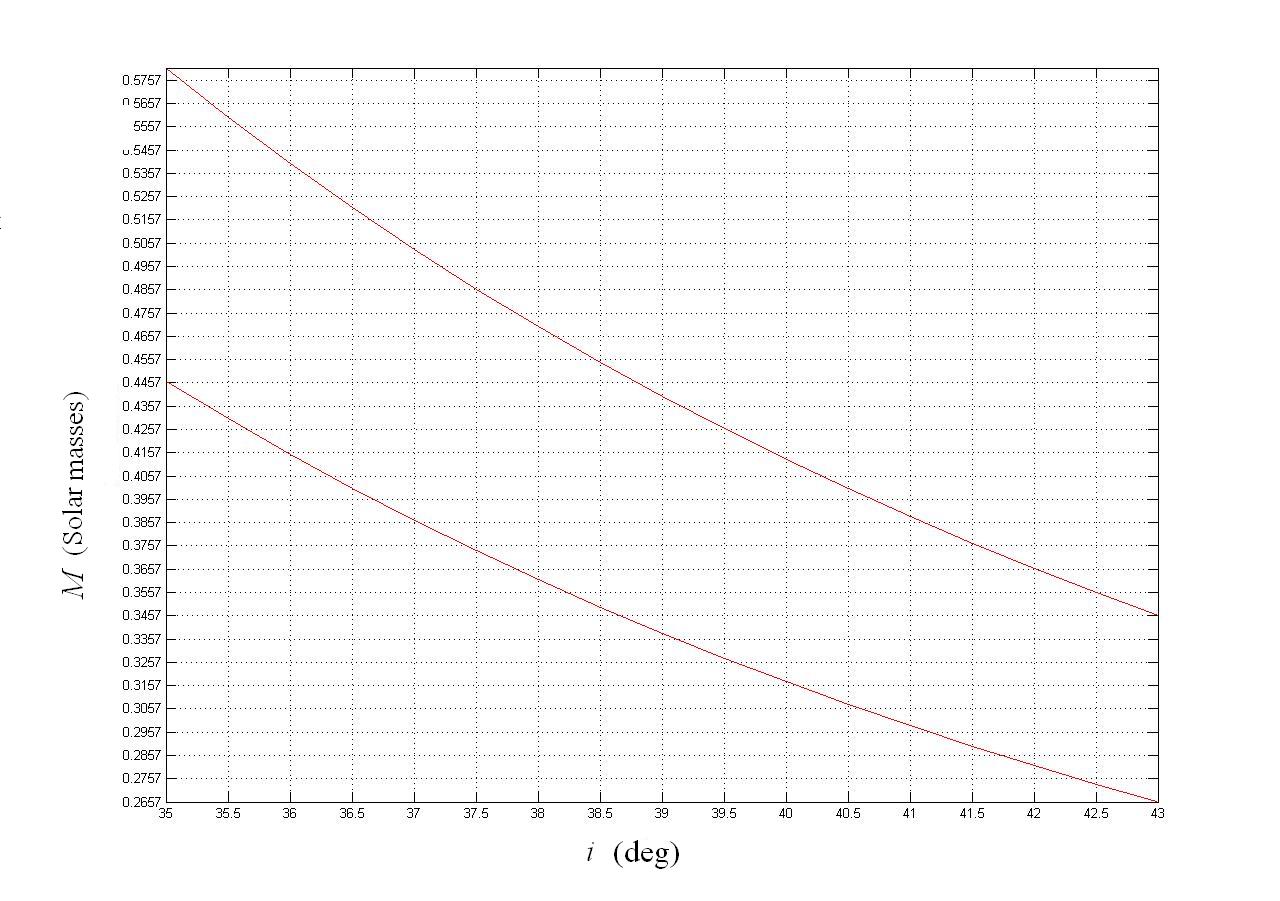}
\end{center}
\caption{\label{MASSAWD}  The allowed values for $M$ (Solar masses), obtained  by modeling the orbital period with the third Kepler law, lie in the region delimited by the upper curve ($M+\delta M$) and the lower ($M-\delta M$) curve.}
\end{figure}

It is interesting to see what are the dynamical constraints on $M$, obtained by modeling the orbital period with the third Kepler law, and compare them to the spectroscopic ones yielding \rfr{massaWD}.
Indeed, from \rfr{pkep}, \rfr{rav} and \rfr{sam} I get
\eqi GM =(1+\gamma)^2\left(\rp{P_{\rm b}}{2\pi}\right)\left(\rp{K_{\rm m}}{\sin i}\right)^3,\lb{erraz}\eqf
used to draw Figure \ref{MASSAWD}  in which the region delimited by $M\pm \delta M$ is displayed as a function of $i$.
Note that the admitted range of mass values for $i=35$ deg is $0.4457 < M <0.5757$, which does not overlap with the spectroscopic one of \rfr{massaWD}. 
\section{Discussion and conclusions}
In this paper I got dynamical constraints on some orbital and physical parameters of the \WD A/B\ binary system consisting of a white dwarf and a brown dwarf as a companion. By using the correction to the third Kepler law due to the quadrupole mass moment $Q$,  I was able to put constraints to the inclination angle $i$ of the system. By discarding those values of $i$ which would yield significant deviations from the third Kepler law associated to unlikely large  $Q$   I obtained $i=39\pm 2$ deg.  The spectroscopically inferred range of admissible values for the mass $M$ of the white dwarf is compatible with our determination of $i$ which also yields an orbital separation of $a=(0.59\pm 0.05)$R$_{\odot}$ and  an equilibrium temperature of the brown dwarf of $T_{\rm eq}=(2087\pm 154)$K.   An incorrect value quoted in \citep{Max06} of $i=35$ deg yields, as a consequence, $a=0.65$R$_{\odot}$ and $T_{\rm eq}\approx 2000$K.

\section*{Appendix: the contribution of the quadrupole mass moment to the orbital period}\label{appendice}
One of the six Keplerian orbital elements in terms of which it is
possible to parameterize the orbital motion in a
binary system is the mean anomaly  $\mathcal{M}$ defined as
$\mathcal{M}\equiv n(t-T_0)$, where $n$ is the mean motion and
$T_0$ is the time of periastron passage. The mean motion $n\equiv
2\pi/ P_{\rm b}$ is inversely proportional to the time elapsed
between two consecutive crossings of the periastron, i.e. the
anomalistic period $P_{\rm b}$. In Newtonian mechanics, for two
point-like bodies, $n$ reduces to the usual Keplerian expression
$n^{(0)}=\sqrt{GM/a^3}$, where $a$ is the semi-major axis of the
relative orbit  and
$M\equiv m_{\rm 1}+m_{\rm 2}$ is the sum of the masses. In
many binary systems, as \WD, the period $P_{\rm b }$ is  accurately
determined in a phenomenological, model-independent way, so that
it accounts for all the dynamical features of the system, not only
those coming from the Newtonian point-like terms, within the
measurement precision.

Here I wish to calculate the contribution of the quadrupole mass moment $Q$ to the orbital period in a more general and accurate way than done in \citep{Iorio07} by retaining  the orbital eccentricity. According to \citet{Iorio07}, the acceleration induced by the quadrupole mass moment $Q$ can be cast into the form
\eqi
A_Q= \rp{3}{2}\rp{GQ}{r^4}.\lb{radialacc}
\eqf

The quadrupole mass term $A_Q$ is small with respect to the
usual Newtonian monopole term $A_0$, so that it can be treated perturbatively. In
order to derive its impact on the orbital period $P_{\rm b}$, let
us consider the Gauss equation for the variation of the mean
anomaly in the case of an entirely radial disturbing acceleration
$A$ \eqi\dert{\mathcal{M}} t=n-\rp{2}{na}A
\left(\rp{r}{a}\right)+\rp{(1-e^2)}{nae}A\cos f,\lb{gauss}\eqf
where $f$ is the true anomaly, reckoned from the periastron. After
inserting $A_Q$ into the right-hand-side of \rfr{gauss}, it must
be evaluated onto the unperturbed Keplerian ellipse \eqi
r=\rp{a(1-e^2)}{1+e\cos f}.\eqf By using \citep{Roy05} \eqi \dert
f{\mathcal{M}} =\left(\rp{a}{r}\right)^2\sqrt{1-e^2},\eqf
\rfr{gauss} yields \eqi\dert f t=\rp{n(1+e\cos f
)^2}{(1-e^2)^{3/2}} \left\{1+\rp{3GQ(1+e\cos f)^3}{n^2
a^5(1-e^2)^3}\left[\rp{\cos f(1+e\cos
f)}{2e}-1\right]\right\}.\lb{grossa}\eqf
The orbital period can
be obtained as \eqi P_{\rm b}\approx
\rp{(1-e^2)^{3/2}}{n}\int_0^{2\pi}\left\{1-\rp{3GQ(1+e\cos
f)^3}{n^2 a^5(1-e^2)^3}\left[\rp{\cos f(1+e\cos
f)}{2e}-1\right]\right\}\rp{df}{(1+e\cos f)^2}.\lb{appro}\eqf
%
From \rfr{appro} it can be obtained \eqi P_{\rm
b}\equiv P^{(0)}+P^{(Q)},\eqf with\footnote{It agrees with the expression of the anomalistic period of a satellite orbiting an oblate planet obtained in \citep{Cap05}: for a direct comparison $Q=-MR^2 J_2$, where $J_2$ is the first even zonal harmonic of the multipolar expansion of the Newtonian part of the gravitational potential of the central body of mass $M$ and equatorial radius $R$.}
\begin{equation}\left\{\begin{array}{lll}
P^{(0)}=2\pi\sqrt{\rp{a^3}{GM}},\\\\
P^{(Q)}=\rp{3\pi Q}{\sqrt{GaM^3(1-e^2)^3}}.\lb{periodoQ}
\end{array}\right.\end{equation}
Solving for $Q$, I get \eqi Q=\rp{ P_{\rm b}\sqrt{GaM^3(1-e^2)^3}
}{3\pi}-\rp{2}{3}Ma^2(1-e^2)^{3/2}.\lb{QU}\eqf

The uncertainty in $Q$ can
 be conservatively assessed by linearly adding the various sources of errors as \eqi \delta Q\leq \delta Q|_a +
\delta Q|_G + \delta Q|_M + \delta Q|_{P_{\rm b}} + \delta
Q|_e,\lb{erroreQ}\eqf with
\begin{equation}\left\{\begin{array}{lll}
\delta Q|_a \leq \left|\rp{P_{\rm b}}{6\pi}\sqrt{\rp{GM^3(1-e^2)^3}{a}}-\rp{4}{3}Ma(1-e^2)^{3/2}\right|\delta a,\\\\
\delta Q|_G \leq \left|\rp{P_{\rm b}}{6\pi}\sqrt{\rp{aM^3(1-e^2)^3}{G}}\right|\delta G,\\\\
\delta Q|_M \leq \left|\rp{P_{\rm
b}}{2\pi}\sqrt{GMa(1-e^2)^3}-\rp{2}{3}a^2(1-e^2)^{3/2}\right|\delta
M,\\\\
\delta Q|_{P_{\rm b}} \leq \left|\rp{\sqrt{GaM^3(1-e^2)^3}
}{3\pi}\right|\delta P_{\rm b},\\\\
\delta Q|_e \leq  \left|\rp{ P_{\rm b}\sqrt{GaM^3(1-e^2)}
}{\pi}-2Ma^2\sqrt{1-e^2}\right|e\delta e.
 \lb{erroriQ}
\end{array}\right.\end{equation}


\end{document}